\documentclass[a4paper,10pt,twoside]{cpc-hepnp}

\usepackage{multicol}
\usepackage{graphicx}
\usepackage{booktabs}
\usepackage{amssymb,bm,mathrsfs,bbm,amscd}
\usepackage[tbtags]{amsmath}
\usepackage{lastpage}
\usepackage{CJK}
\usepackage{color}

\begin{document}
\begin{CJK*}{GBK}{song}

\fancyhead[c]{\small Chinese Physics C~~~Vol. cc, No. c (cccc)
cccccc} \fancyfoot[C]{\small cccccc-\thepage}


\title{LOCV calculation of the equation of state {and} properties of rapidly rotating neutron stars}

\author{%
      A.H. Farajian%
\quad M. Bigdeli$^{1)}$\email{m\underline\ \ bigdeli@znu.ac.ir}%
\quad S. Belbasi
}
\maketitle

\address{%
 Department of Physics, University of Zanjan,\\
Zanjan, 45371-38791, Iran}

\begin{abstract}
In this paper, we have investigated the structural properties of
rotating neutron stars using the numerical RNS code and
equations of state which have been calculated within the lowest
order constrained variational (LOCV) approach. In order to calculate the
equation of state of nuclear matter, we have used UV$_{14}$ $+$TNI
and AV$_{18}$  potentials. We have computed  the maximum
mass of the neutron star and the corresponding equatorial radius
at different angular velocities. We have also computed the
structural properties of Keplerian rotating neutron stars for the
maximum mass configuration, $M_{K}$, $R_{K}$, $f_{K}$ and
$j_{max}$.
\end{abstract}




\begin{multicols}{2}

\section{Introduction}

All existing studies indicate that observed neutron
stars, such as millisecond pulsars~(MSPs), are rotating.
Recently, many MSPs have been discovered. One
of the most rapidly rotating neutron stars is pulsar PSR
J1748-2446ad, which has rotational frequency 716 Hz~\cite{hessel}. The rotational frequency $f$, which can be
directly measured, affects the global attributes of  neutron
stars, specifically, maximum mass, radius, spin parameter and
total moment of inertia~\cite{hart, niko, salga,cook,Haensel}. The
maximum mass increases with rotation due to the rotational energy
and there are even super-massive sequences~\cite{salgb}.
So far, there have been a  large number of mass and radii measurements. The
accurate measurement of mass for about 35 neutron stars lies in
the wide range of $M\sim1.17-2.0~M_{\odot}$ and the radii of more
than a dozen neutron stars lies in the range $R\sim9.9-11.2$
km~\cite{ozel}. Two well-measured massive neutron stars are MSPs
in binary systems, PSR J1614-2230 with mass $M=1.928\pm0.017
M_{\odot}$~\cite{dem}, and PSR J0348+0432, with mass $M=2.01\pm0.04 M_{\odot}$ \cite{ant}. These massive neutron stars
require the equation of state~(EOS) of the system to be rather
stiff.
Present radius determinations are model dependent and subject to
large uncertainties. However, some current and planned projects,
such as NICER \footnote{https://www.nasa.gov/nicer} are trying to
determine the radii more precisely. Theoretically, the EOSs have been
applied to determine neutron star properties which should be in
agreement with the precise observations.

Another important characteristic quantity for compact stars is the dimensionless spin parameter $j\equiv cJ/GM^2 $,
where $J$ is angular momentum and $M$ is gravitational mass.
The astrophysical estimations and implications of $j$ for
different astronomical objects have been considered  by several
authors, e.g. Refs.~\cite{duez,piran,shibata,torok1,torok2, kato}.
T\"{o}r\"{o}k \textit{et al.} have investigated the \textit{mass
vs. spin parameter} relationship $M(j)=M_0[1+k(j+j^2)]$ for the
Z-source Circinus X-1~\cite{torok1} and atoll source $4U 1636-53$
\cite{torok2}. Kato
\textit{et al.} have shown that a description of the observed
correlations of Circinus X-1 requires adopting $M=1.5- 2.0~M_{\odot}$
as the mass of the central star in Circinus X-1 and $j\sim 0.8$
for the dimensionless spin parameter~\cite{kato}.
Recently, this parameter has been studied in detail for uniformly
rotating compact stars by Lo and Lin~\cite{1}. They have
discussed that the spin parameter plays an important role in
understanding the observed quasi-periodic oscillations (QPOs) in
disk-accreting compact-star systems. They have shown that the
maximum value of the spin parameter, $j_{max}$ (spin parameter of a
neutron star rotating at the Keplerian frequency), depends on the
composition of compact stars. Their results indicate that the
value of $j_{max}$ has an upper bound about $j_{max} \sim 0.7$ for
traditional neutron stars; and it is independent of the EOS and
also insensitive to the mass of the star for $M\geq1 M_\odot$
\cite{1}. Their results also indicate that there is no universal
upper bound for the spin parameter of quark stars simulated by the MIT
bag model and it can be larger than unity~($j_{max}> 1$).
A different point of view has been followed by Qi et al. \cite{2};
they have found that the crust structure of compact stars is essential
to determine the maximum value of the spin parameter. They have concluded
that when the whole crust EOS is not considered, $j_{max}$ of
compact stars can be larger than $0.7$ but also less than $1$ for
traditional and hyperonic neutron stars and also for hybrid stars,
whereas the role of the crust in the total mass of the compact
star is negligible. In this paper, we show that only the outer
crust structure could play the same roles, see Section~3. Qi et al.
also have constructed a universal formula for spin parameter versus
frequency, $j=0.48(f/f_{k})^{3}-0.42(f/f_{k})^{2}+0.63(f/f_{k})$,
for different kinds of compact stars.

In this study, we have investigated the structural properties
of rapidly rotating neutron stars with and without outer crust structures. Here we have used EOS for the liquid
core of the neutron star which have been calculated within the lowest
order constrained variational~(LOCV) method with UV$_{14}$ $+$TNI~\cite{pand} and
AV$_{18}$~\cite{wiring} potentials. Previously, we used these EOS to determine the core-crust transition parameters and global attributes of core and crust for neutron stars~\cite{bigelyas}.

\section{Neutron star matter equation of state}\label{S2}
We have employed the EOS for neutron star matter by
describing the neutron star's outer crust, inner crust and the
liquid core. For the inner crust, we use the EOS which is
calculated by Douchin and Haensel \cite{dh}, and for the outer
crust, the Baym-Pethick-Sutherland EOS \cite{byme} is used. In
the case of the neutron star core, we assume a charged neutral
infinite system which is a mixture of leptons and interacting
nucleons. The energy density of this system can be obtained as
follows,
\begin{equation}
\varepsilon=\varepsilon _{N}+\varepsilon _{l},
\end{equation}
where $\varepsilon_{N}$~($\varepsilon_{l}$) is the energy density
of nucleons~(leptons). The energy density of leptons, which are
considered as a noninteracting Fermi gas, is given by,
%
\begin{eqnarray}
       \varepsilon_{lep}&=&\sum_{l=e,\ \mu}\ \sum_{k\leq k_l^{F}}
       (m_{l}^2 c^4+\hbar^2 c^2 k^2)^{1/2} \ .
 \end{eqnarray}
In this equation, $k_l^{F}=(3\pi^2\rho_l)^{1/3}$ is
the Fermi momentum of leptons.
The nucleon contribution of energy density is given by,
\begin{equation}
\varepsilon _{N}=\rho(E_{nucl}+m_Nc^2),
\end{equation}
where $E_{nucl}$ is the total energy per particle of asymmetric
nuclear matter and $\rho$ is the total number density,
%
$$\rho=\rho_{p}+\rho_{n}.$$
 %
Here, $\rho_{n}$ and $\rho_{p}$  are number density of neutrons and
protons respectively. \\
\\

%
\begin{center}
\tabcaption{ \label{tab}  Saturation density and corresponding values of energy per particle, incompressibility and symmetry energy of symmetric nuclear matter. Here
$\rho_s$ is given in fm$^{-3}$ and energy parameters are in MeV.}
\footnotesize
\begin{tabular*}{70mm} {@{\extracolsep{\fill}}c|cccc}
\toprule Potential& $\rho_s$& $E_0$ &$K_0$ & $S_0$ \\
  \hline
  UV$_{14}$ $+$ TNI\hphantom{00} &0.17 &-16.86 & 261&31.27 \\
   AV$_{18}$ \hphantom{00}        &0.31 &-18.47&301&36.24 \\
 \bottomrule
  \end{tabular*}
\end{center}
%
%
The $\beta$-equilibrium conditions and
charge neutrality of neutron star matter impose the following
coupled constraints on our calculations,
\begin{eqnarray}\label{fbeta}
\mu_e=\mu_{\mu}&=&\mu_n-\mu_p
\end{eqnarray}
\begin{equation}\label{neutrality}
\rho_p = \rho_e + \rho_{\mu}.
\end{equation}
We find the abundance of the particles by solving these coupled
equations and calculate the total energy and the EOS
of the neutron star matter.

In the following, we determine the energy per particle of asymmetric nuclear matter,
$E_{nucl}$, in more detail by using the LOCV method.
%
%
In our formalism, the energy per particle is written in terms of correlation function, $f$, and its derivatives; and approximately
given up to the two-body term as the following form~\cite{clark},
 \begin{eqnarray}\label{tener}
           E_{nucl}([f])=\frac{1}{A}\frac{\langle\psi|H|\psi\rangle}
           {\langle\psi|\psi\rangle}
           =\frac{1}{A}\sum _{\tau=n,p}{\sum _{k\leq{k_\tau^F}}
               \frac{\hbar^{2}{k^2}}{2m_\tau}} \nonumber \\+\frac{1}{2A}\sum_{ij} \langle ij\left| \nu(12)\right|
    ij\rangle_a,
 \end{eqnarray}
where $\psi=\cal{F}\phi$ is a trial many-body wave function.
Here $\phi$ is the Slater determinant of wave function of $A$
independent nucleons and ${\cal F}={\cal S}\prod _{i>j}f(ij)$ (${\cal S}$ is a symmetrizing operator) is a Jastrow form of
A-body correlation operator.
In the above equation,
$k_\tau^F=(3\pi^2\rho_\tau)^{1/3} $ is the Fermi momentum of nucleons and $\nu(12)$ is the effective potential, which is given by,
\begin{eqnarray}
 \nu(12)=-\frac{\hbar^{2}}{2m}[f(12),[\nabla
_{12}^{2},f(12)]]+f(12)V(12)f(12).
\end{eqnarray}
Here, $f(12)$ and $V(12)$ are the two-body correlation and
potential, respectively. In our calculations, we used the UV$_{14}$ $+$TNI and $AV_{18}$ two-body
potentials.
%
\begin{center}
\includegraphics[width=10cm]{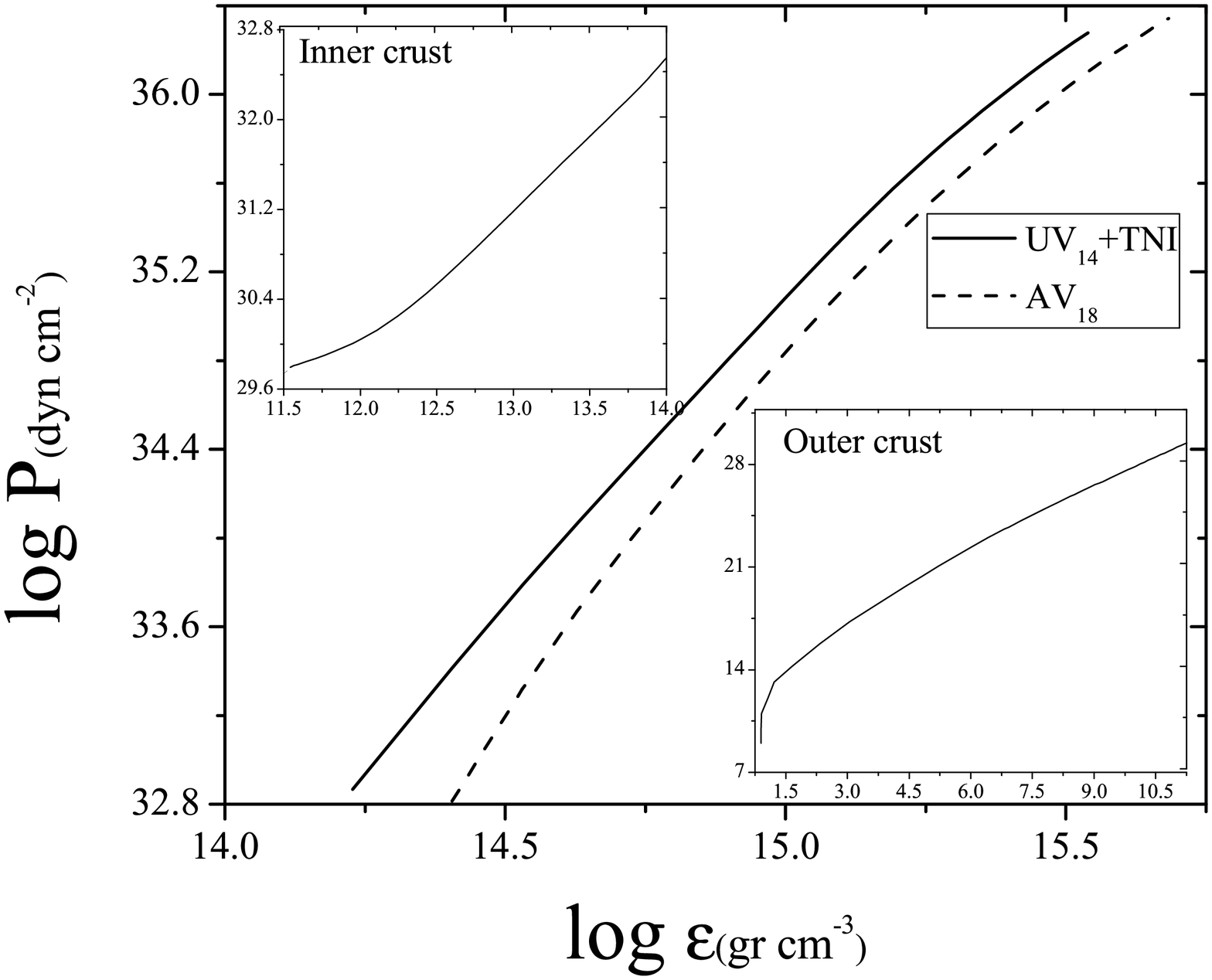}
\figcaption{The EOS of neutron star matter for the UV$_{14}$ + TNI and $AV_{18}$ potentials. The EOS of the outer and inner crust are also shown.} \label{eos}
\end{center}
%

In this formalism, the correlation function is considered as different forms~\cite{OBI3}, and  calculated by numerically solving of set of coupled and uncoupled Euler-Lagrange differential equations~\cite{BM98}. These differential equations are a result of functional minimization of the two-body cluster energy with respect to the correlation functions variation. For more details see Refs.~\cite{BM98,MB98,big1,bordbig6}. 

{A summary of our results for bulk properties of symmetric nuclear matter for the UV$_{14}$ $+$TNI and $AV_{18}$ potentials are given in Table~\ref{tab}. In this table, we have given the saturation density $\rho_s$, and the corresponding values of energy per particle $E_0$, incompressibility $K_0$, and nuclear symmetry energy $S_0$. The calculated saturation properties of symmetric nuclear matter are
in excellent agreement with the experimental data~\cite{haus} for the UV$_{14}$ $+$TNI potential.}

The pressure of neutron star matter can be calculated by the following relation, 
\begin{eqnarray}
P&=&\rho\frac{\partial \varepsilon}{\partial \rho}- \varepsilon.
\end{eqnarray}
In Fig.~\ref{eos}, we have plotted the pressure of neutron star matter at the core of the star for the mentioned
potentials versus total energy density. In this figure we  also show the EOS for outer and inner crust.
It is seen that the UV$_{14}$ $+$TNI potential leads to a stiffer EOS.

\begin{center}
\includegraphics[width=9cm]{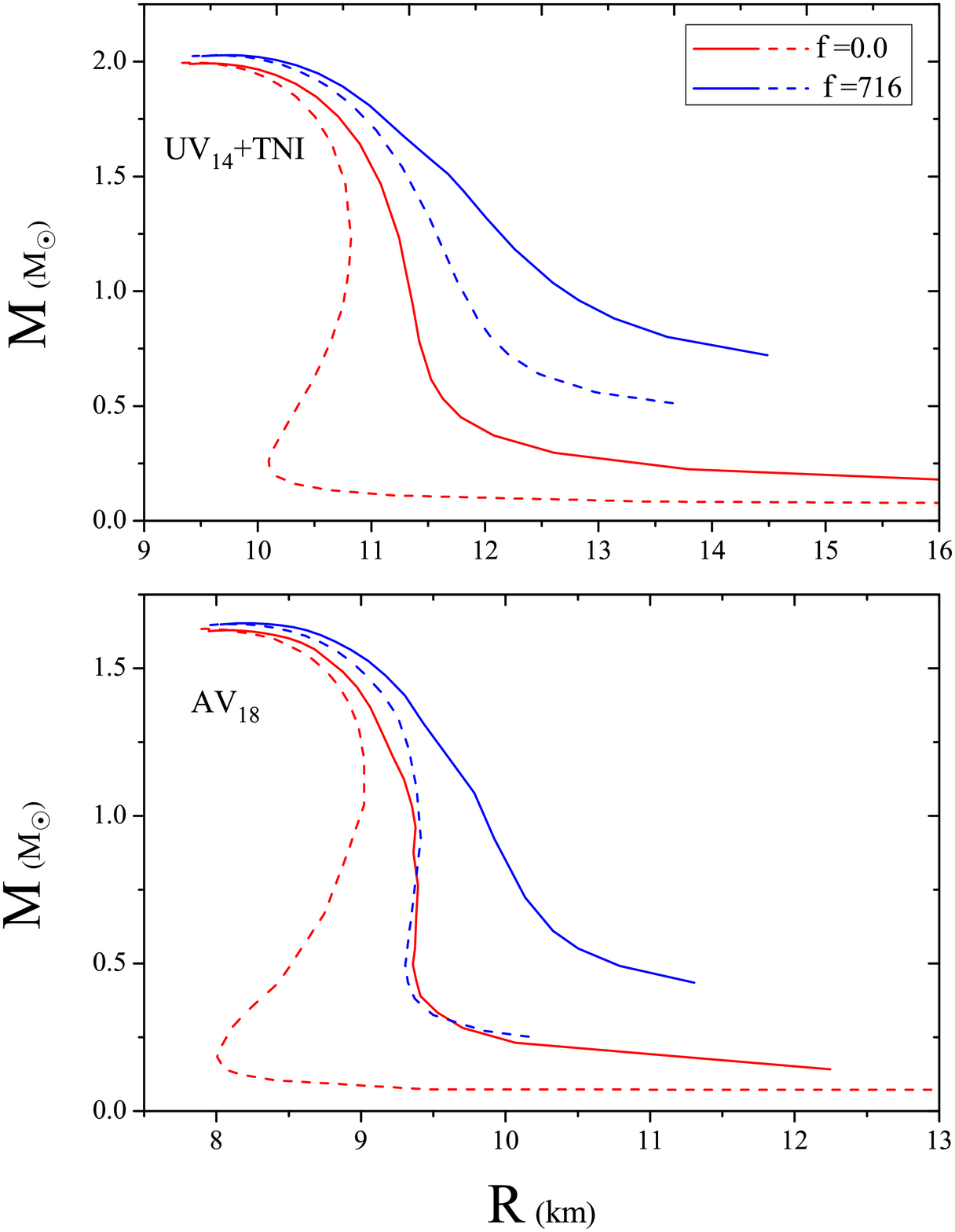}
\figcaption{The gravitational mass~($M$) versus
circumferential radius~($R$) for non-rotating and rotating neutron
star with the UV$_{14}$ +TNI and AV$_{18}$ potentials. The
frequency~($f$) is given in Hz. The solid (dashed) curve shows the
result for neutron stars including~(excluding) the outer crust
structure.} \label{mr}
\end{center}
%
\section{Results and discussion}
\label{sec:maths}

We now proceed to show our results for rotating neutron stars. We
make use of the numerical RNS
code~(http://www.gravity.phys.uwm.edu/rns/), which  integrates
the Einstein field equations for a rapidly rotating neutron star given a perfect fluid
EOS \cite{ster}. In Fig.~\ref{mr}, we show
the gravitational mass versus (circumferential) radius for two
different microscopic EOS at fixed frequency $f=0$ and $f=716$
Hz.
The solid (dashed) curve shows the result for neutron stars including~(excluding) the outer crust structure. Clearly, the
inclusion of the outer crust has no considerable effect on the
maximum mass and corresponding radius of the neutron star.
However, the global structure of the neutron star is sensitive to
its angular velocity, and the maximum mass increases by increasing
 the rotation velocity.
%
\begin{center}
\includegraphics[width=10cm]{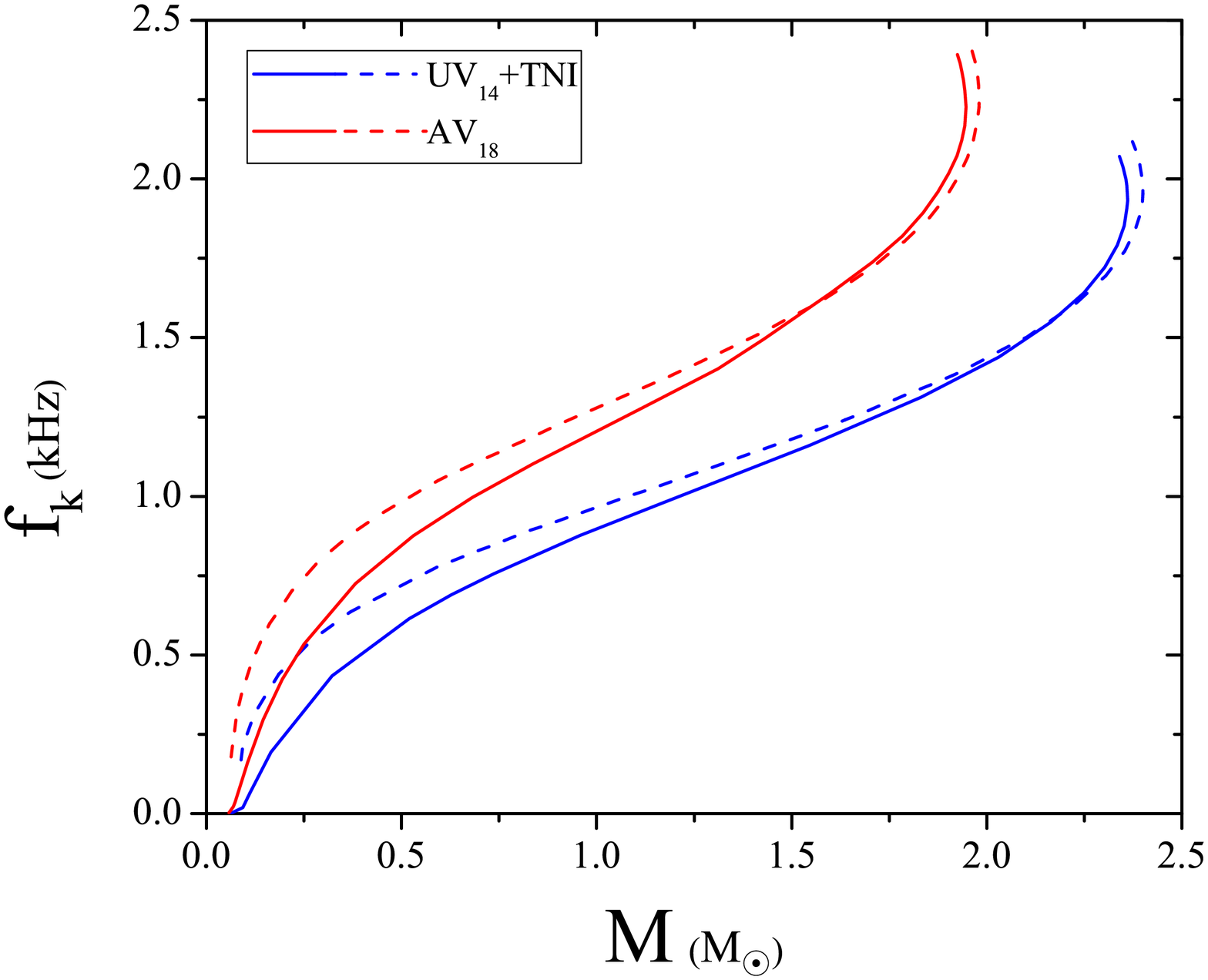}
\figcaption{The variation of the Keplerian
frequency~($f_K$) with gravitational mass~$M$ for neutron stars
with~(solid curve) and without~(dashed curve) outer crust
structure.}
  \label{fkm}
\end{center}
From this figure, one can compare the results of the EOS derived using the UV$_{14}$+TNI and AV$_{18}$ potentials. At a frequency of
$f=716$ Hz, which corresponds to the spin period $P\approx1.39$
ms, by applying the
AV$_{18}$ potential, we get $M_{max}/M_\odot \simeq 1.653$ ($\simeq
1.649$) for a neutron star with~(without) outer crust
structure. Using the UV$_{14}$ +TNI leads to larger stellar mass and
radius in comparison with the AV$_{18}$ potential, and  we obtain $M_{max}/M_{\odot} \simeq 2.0278$ ($\simeq 2.0275$)
with the UV$_{14}$ +TNI potential.
This is in good agreement with the results obtained by
observations for the millisecond pulsar PSR J0348+0432,
$M=2.01\pm0.04 M_{\odot}$ \cite{ant}. {However, this pulsar rotates with the lower frequency of $\simeq 25$ Hz. This does not affect the good comparison, because in this frequency range the maximum mass has a little variance with the rotation (see Fig.~\ref{mr} and Table~\ref{tab1}). }

Another crucial parameter that can be used to describe
rotating neutron stars is the Keplerian frequency, $f_k$, the maximum
value of frequency.  We have plotted Keplerian frequencies versus
gravitational masses in Fig.~\ref{fkm}. It is seen that $f_{k}$
depends on the  EOS models presented here. From Fig.~\ref{fkm},
for the case of the UV$_{14}$ +TNI potential, we find that the value
of the Keplerian mass corresponding to our calculated frequency,
$f_k \simeq 1.93$ kHz ($f_k\simeq 1.96$ kHz) is about $M_k\simeq
2.36 M_\odot $ ($M_k\simeq 2.40 M_\odot$) for a neutron star with~(without)
outer crust structure. For the case of the AV$_{18}$ potential, we
find $M_k\simeq 1.95 M_\odot $ ($M_k\simeq 1.98 M_\odot$)
corresponding to $f_k \simeq 2.23$ kHz ($f_k\simeq 2.24$ kHz). It
is seen that the Keplerian mass and frequency for a neutron star with outer crust
are a little lower than those of a neutron star without outer crust.
%
\begin{center}
\includegraphics[width=10cm]{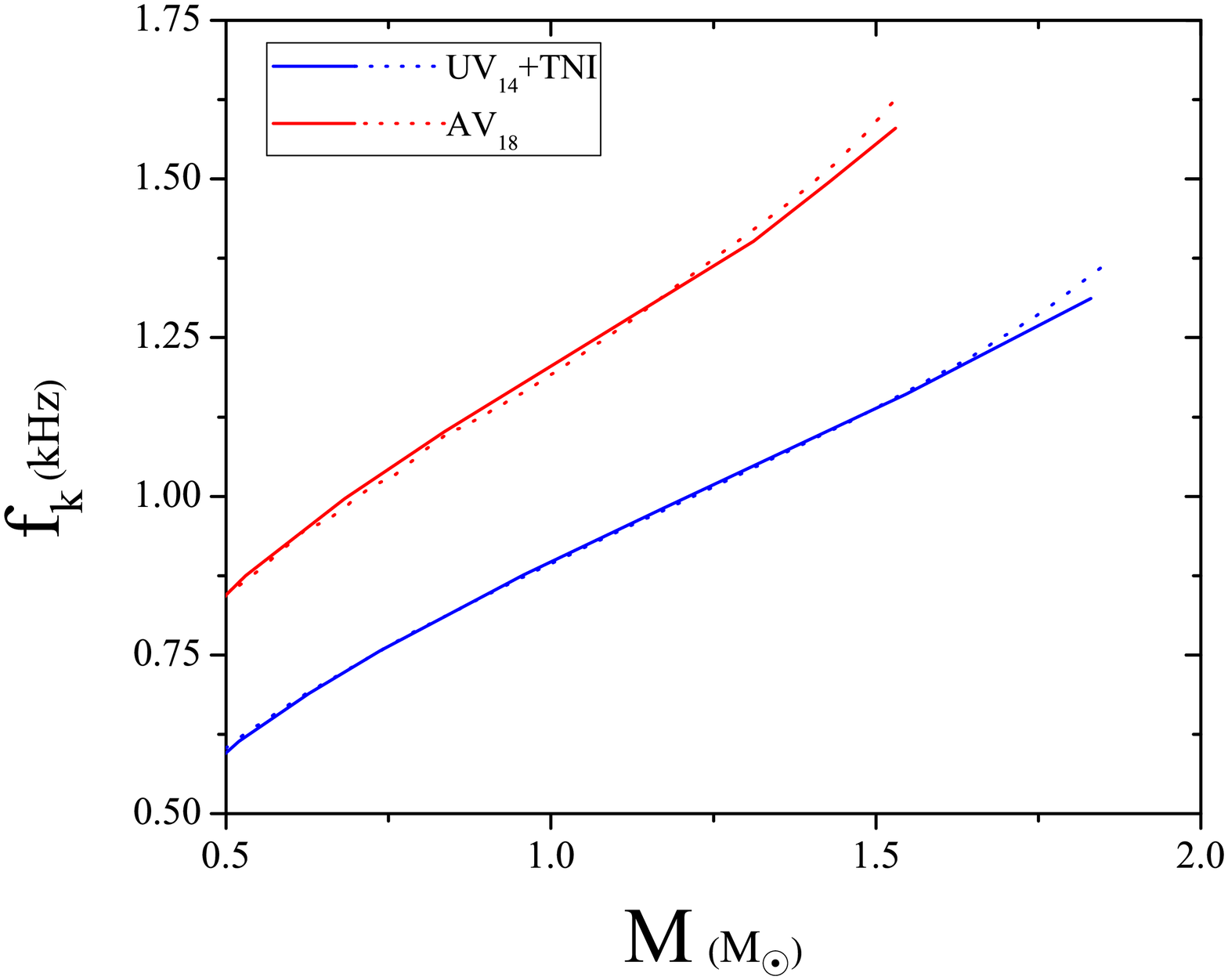}
\figcaption{The variation of the Keplerian
	frequency~($f_K$) with gravitational mass~$M$ for neutron stars  for precise values
of Keplerian frequency~(solid curve) and those given by
Eq.~(\ref{fk})~(dotted curve).}
  \label{fkm2}
\end{center}
%
We have also calculated Keplerian frequency using the fit formula proposed
by Haensel et al.~\cite{Haensel},
\begin{eqnarray}\label{fk}
               f_K&=& 1.08~kHz\left(\frac{M}{M_\odot}\right)^{1/2}\left(\frac{R}{10~km}\right)^{-3/2},
            \end{eqnarray}
where $0.5 M_{\odot}\leq M \leq 0.9 M^{stat}_{max}$, $M^{stat}_{max}$ is the maximum mass of the non-rotating~(static) configuration and $R$ is the corresponding radius. The results are shown in Fig.~\ref{fkm2}. As can be seen from this figure,
there is a good agreement between the precise values and those
calculated using the above equation, especially for the UV$_{14}$ +TNI
potential.

In the following, we discuss the relation between maximum
mass and frequency in more detail. In Fig.~\ref{mf}, we
present the maximum mass in units of Keplerian mass,
$M_{max}(M_K)$, as a function of stellar frequency in units of
Keplerian frequency, $f(f_K)$.
This figure shows that for both EOS employed in the present work,
$M_{max}(M_K)$ displays a similar behavior versus $f(f_K)$ and,
nearly, does not depend on the EOS. According to this behavior, we
find $$0.835 {M_K} \lesssim M_{max}\leq 1.0 {M_K}.$$ In other
words, the maximum mass in the Keplerian configuration increases about
$20\% $ compared to the maximum mass of non-rotating configurations. This
result is in agreement with those obtained by the universal relation
$M_k\simeq(1.203\pm0.022)M^{stat}_{max}$  proposed by Breu and
Rezzola~\cite{breu}.
%
\begin{center}
\includegraphics[width=10cm]{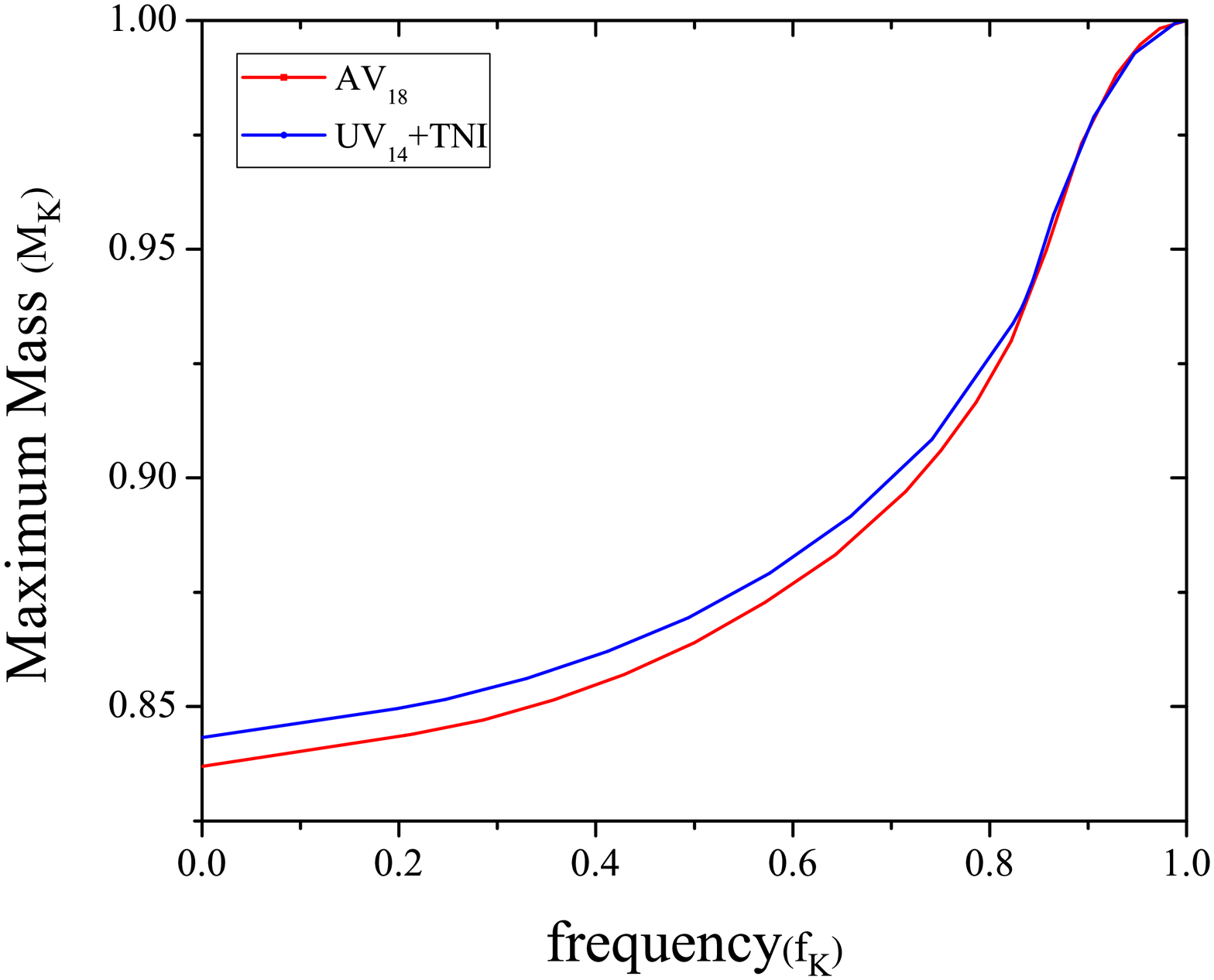}
\figcaption{The maximum mass versus frequency for different equations of state.
The maximum mass and frequency are given in units of Keplerian mass and frequency, respectively.}
\label{mf}
\end{center}

Now, we focus on the treatment of the dimensionless spin
parameter $j$, for rotating neutron stars. Here, we would like to
consider the influence of the outer crust structure on the spin
parameter at Keplerian frequency, i.e. maximum spin parameter,
$j_{max}$. In order to achieve this, we shown the maximal
spin parameter, $j_{max}$, as a function of gravitational mass in
Fig.~\ref{jm}. As can be seen from this figure, the maximal spin
parameter of the rotating neutron star displays different
behaviors when we either include or exclude the outer crust
structure. It is seen that $j_{max}$ for NSs with the outer crust
structure lying in the narrow range $\sim$ (0.64 \-- 0.7) for $M
\geq 0.5 M_{\odot}$. Therefore, we see that our result for the
upper limit of $j_{max}$($\leq 0.7$) is in agreement with those
reported earlier \cite{1,2} for traditional neutron stars, while,
for the neutron star with only inner crust structure  $j_{max}$ is larger
than 0.7 and this value is the lower limit of $j_{max}$($\geq
0.7$).
This shows that, in spite of the role of outer crust structure
in the maximal mass, its role in maximal spin parameter is important. It is worth noting that the similar results have been concluded in the work by Qi et al.,
but they have considered the whole crust structure in calculating the maximum value of the spin parameter~\cite{2}.

Finally, we have investigated the spin parameter, $j$, of slow
rotating neutron stars. In Fig.~\ref{jf}, we plot the spin
parameter $j$ as a function of the rotational frequency
normalized to Keplerian frequency, $f/f_{k}$, for using the
UV$_{14}$+TNI at different values of baryonic mass of neutron
star, $M_{b}/M_{\odot}=1, 1.5, 2$.
%
\begin{center}
\includegraphics[width=10cm]{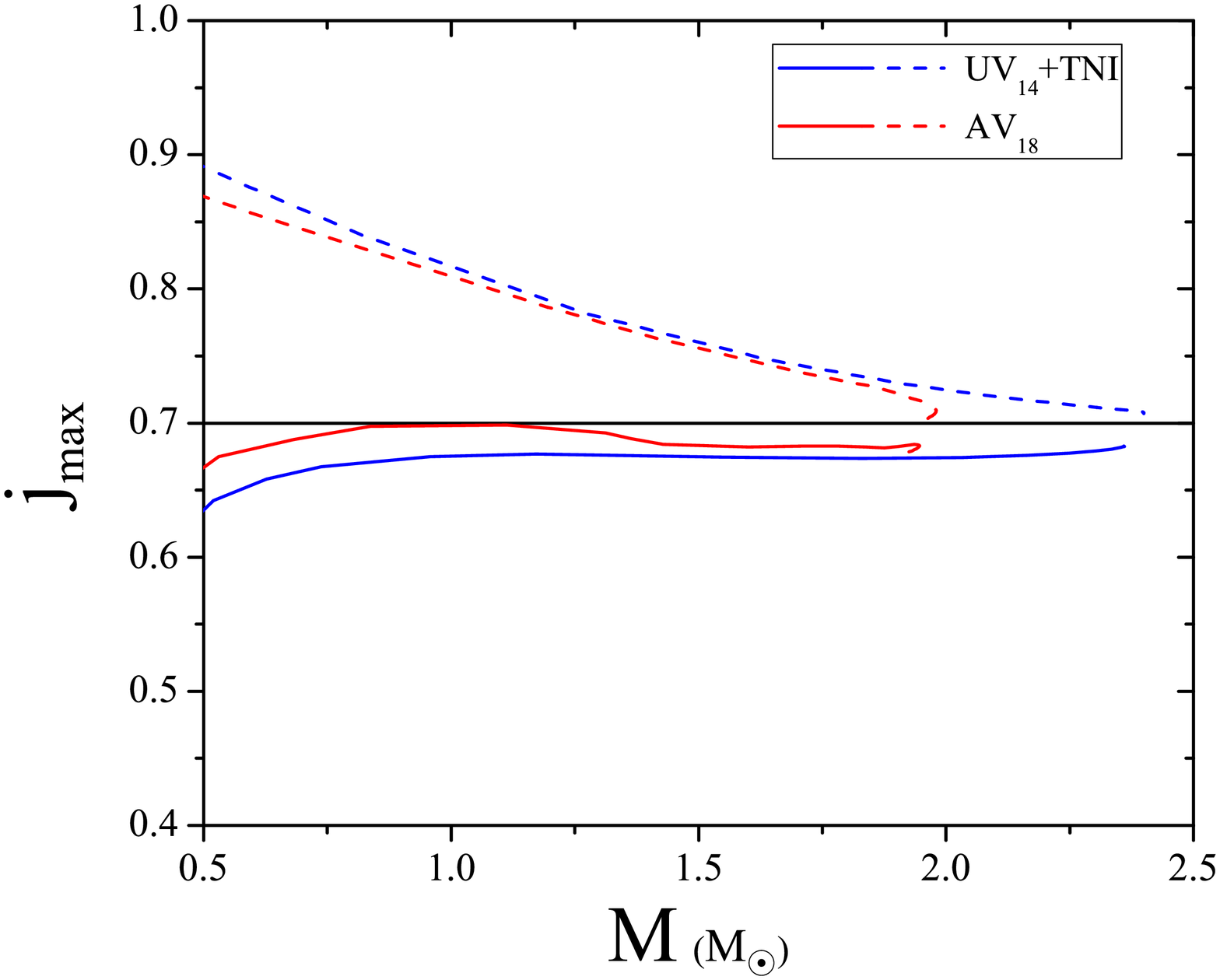}
\figcaption{The variation of the maximum spin parameter~($j_{max})$ with gravitational mass~$M$ for neutron stars
	with~(solid curve) and without~(dashed curve) outer crust
	structure. }
  \label{jm}
\end{center}
%
It is seen that for each fixed
frequency, the curves are essentially independent of mass
sequence. A unified relationship could be fitted approximately by
the formula $j=0.16(f/f_{k})^{3}-0.1(f/f_{k})^{2}+0.612(f/f_{k})$,
as denoted by the circles. We also show the result of the
universal formula
$j=0.48(f/f_{k})^{3}-0.42(f/f_{k})^{2}+0.63(f/f_{k})$, which has
been suggested in Ref.~\cite{2}, with squares, for comparison.
%
\begin{center}
\includegraphics[width=10cm]{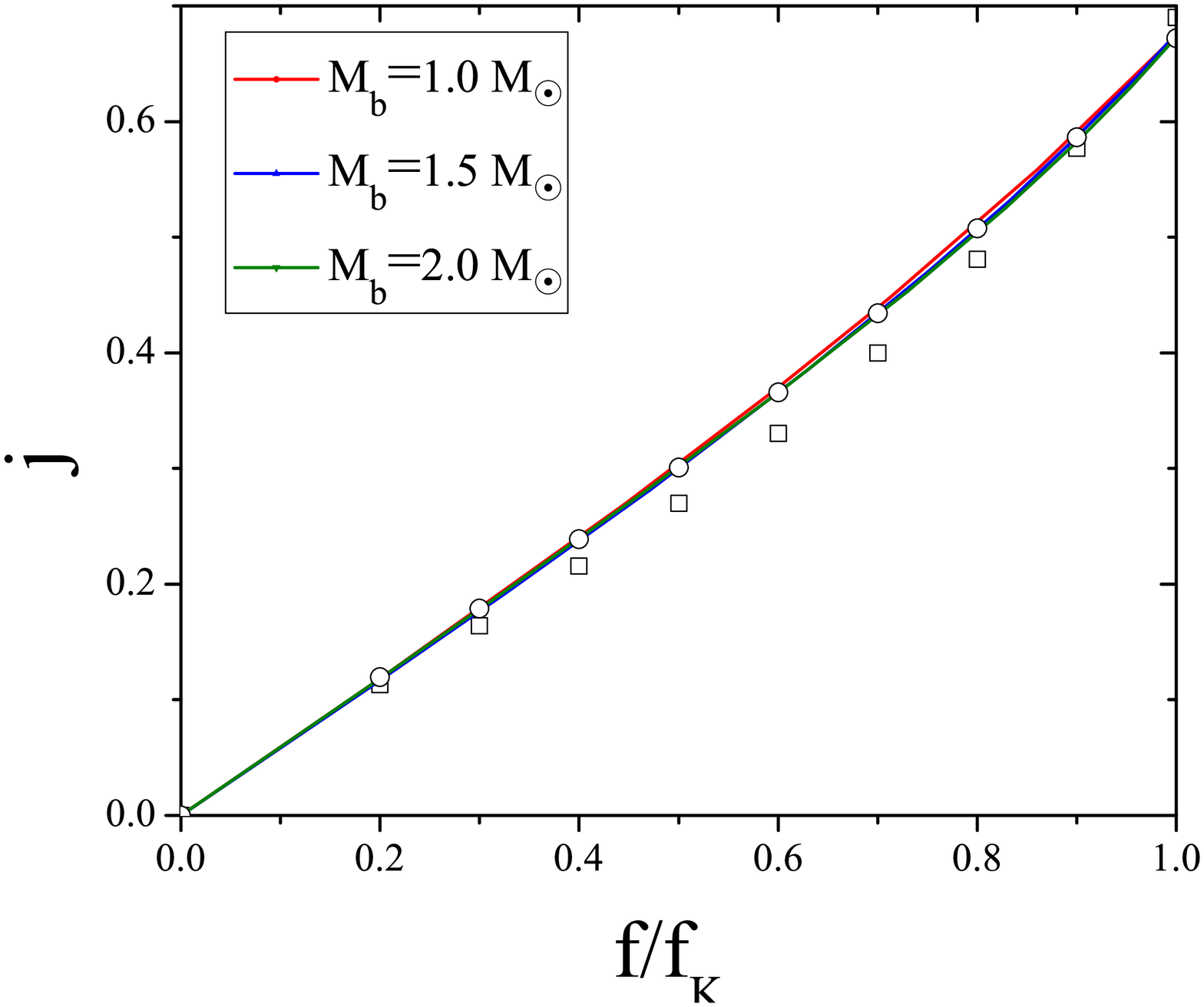}
\figcaption{ The dimensionless spin parameter ($j$) as a function of the
rotational frequency normalized to Keplerian frequency~($f/f_K$) for the UV$_{14}$+TNI potential. The circles show our fitted formula, and squares that from Ref.~\cite{2}.}
  \label{jf}
\end{center}

A summary of our results for the structural
properties of rotating neutron stars with and without outer crust
predicted from different EOS is given in
Table~\ref{tab1}. This table also includes the maximum mass and
corresponding equatorial radius for neutron stars at $f=0,$ and
$716$ Hz, as well as the structural properties of
Keplerian rotating neutron stars for the maximum mass
configuration, $M_{K}$, $R_{K}$, $f_{K}$ and $j_{max}$.

\section{Summary and Conclusions}
In this work, we have calculated the structural properties of
rotating neutron stars with and without outer crust structures.
Here we have employed lowest order constrained variational
approach and used the UV$_{14}$ $+$TNI and AV$_{18}$  potentials
to compute the EOS of nuclear matter. We have
computed maximum mass and corresponding equatorial radius
at fixed frequency $f=0$ and $f=716$ Hz. We have also computed the
structural properties of Keplerian rotating neutron stars for maximum
mass configuration, $M_{K}$, $R_{K}$, $f_{K}$ and $j_{max}$.

A summary of our results for the structural
properties of rotating neutron stars with and without outer crust
predicted from different EOS is given in
Table~\ref{tab1}. Our results show that the maximal spin parameter, $j_{max}$, lies in the narrow range $\sim$ (0.64
\-- 0.7) for $M \geq 0.5 M_{\odot}$ for the EOS considered.
In the case of slow rotating neutron stars, we have suggested a
unified relationship for the spin parameter
$j=0.16(f/f_{k})^{3}-0.1(f/f_{k})^{2}+0.612(f/f_{k})$ which is
essentially independent of mass sequence. Finally,  our results in the Keplerian configuration
are in very good agreement with those of other studies.
\end{multicols}
\begin{center}
\tabcaption{ \label{tab1}  Maximum mass and corresponding equatorial radius for neutron stars at $f=0,\ 716$ Hz. The structural properties of Keplerian rotating neutron stars for maximum mass configuration, $M_{K}$, $R_{K}$, $f_{K}$ and $j_{max}$  are also given. The gravitational mass is given in solar masses ($M_\odot$), $R$ is in km and $f_{K}$ in kHz. The quantities in parenthesis show the results of our calculation for neutron stars without outer crust structure.}
\footnotesize
\begin{tabular*}{180mm} {@{\extracolsep{\fill}}c|cc|cc|cccc}
\toprule Potential& $M_{f=0}$ &  $R_{f=0}$ &  $M_{f=716}$ &  $R_{f=716}$ &$M_{K}$& $R_{K}$ &$f_{K}$ & $j_{max}$ \\
  \hline
  UV$_{14}$ $+$ TNI\hphantom{00}&1.99(1.99) &9.67(9.56)&2.027(2.027)& 9.8(9.7) &2.36(2.40) &12.59(12.59) &1.93(1.96) &0.682(0.707) \\
   AV$_{18}$ \hphantom{00}       &1.63(1.63) &8.09(7.92)&1.653(1.649)& 8.22(8.11) &1.95(1.98) &10.77(10.82)&2.23(2.24)&0.683(0.71) \\
 \bottomrule
  \end{tabular*}
\end{center}

\vspace{20mm}

\acknowledgments{We wish to thank the University of Zanjan Research Councils.}

\vspace{10mm}

\vspace{-1mm}
\centerline{\rule{80mm}{0.1pt}}
\vspace{2mm}

\begin{multicols}{2}

\end{multicols}

\clearpage

\end{CJK*}

\begin{thebibliography}{90}

\vspace{3mm}

\bibitem{hessel} J. W. T. Hessels, et al., { Science}, {\bf 311}: 1901 (2006)
\bibitem{hart}  J. B. Hartle, and K. S. Thorne, { ApJ}, {\bf 153}: 807 (1968)
\bibitem{niko}  N. Stergioulas, and J.L. Friedman, { ApJ}, {\bf 444}: 306 (1995)
\bibitem{salga}  M. Salgado, S. Bonazzola, E. Gourgoulhon, and P. Haensel, { Astron. Astrophys}, {\bf 108}: 455 (1994)
\bibitem{cook}  G. B. Cook,  S. L. Shapiro, and S. A. Teukolsky, { ApJ}, {\bf 424}:  823 (1994)
\bibitem{Haensel}  P. Haensel, J. L. Zdunik, M. Bejger, and   J. M. Lattimer, { Astron. Astrophys}, {\bf 502}: 605 (2009)
\bibitem{salgb} M. Salgado, S. Bonazzola, E. Gourgoulhon, and P. Haensel, { Astron. Astrophys}, {\bf 291}: 155 (1994)
\bibitem{ozel} F. \"{O}zel, and P. Freire, { Annu. Rev. Astron. Astrophys}, {\bf 54}: 401 (2016)
{\bibitem{dem}P.B. Demorest, T. Pennucci, S.M. Ransom, M.S.E. Roberts, and J.W.T. Hessels, {Nature}, {\bf 467}: 1081 (2010)}
\bibitem{ant}  J. Antoniadis, et al., { Science}, {\bf 340}: 448 (2013)
\bibitem{duez}  M. D. Duez, S. L. Shapiro, and H. J. Yo, { Phys. Rev. D}, {\bf 69}: 104016 (2004)
\bibitem{kato}  S. Kato, { Publ. Astron. Soc. Japan}, {\bf 60}: 889 (2008)
\bibitem{piran} T. Piran, { Rev. Mod. Phys}, {\bf 76}: 1143 (2005)
\bibitem{shibata}  M. Shibata, Phys. Rev. D, {\bf 67}:  024033 (2003);  M. Shibata,, { ApJ}, {\bf 595}: 992 (2003)
\bibitem{torok1} G. T\"{o}r\"{o}k, P. Bakala, E. Sr\'{a}mkov\'{a}, Z. Stuchlik, and M. Urbanec, { ApJ}, {\bf 714}: 748 (2010)
\bibitem{torok2} G. T\"{o}r\"{o}k, P. Bakala, E. Sr\'{a}mkov\'{a}, Z. Stuchlik, and M. Urbanec, and K. Goluchov\'{a}, {ApJ}, {\bf 760}: 138 (2012)
\bibitem{1}  K. W. Lo, and L. M. Lin, ApJ, {\bf 728}: 12 (2011)
\bibitem{2} B. Qi, N. B. Zhang, B. Y. Sun, S. Y. Wang, and J. H. Gao, { RAA}, {\bf 16}, No. 4  (2016)
\bibitem{pand} I. E. Lagaris, and V. R. Pandharipande, {Nucl. Phys. A}, {\bf 359}: 349 (1981)
\bibitem{wiring}  R. B. Wiringa, V. G. J. Stoks, and R. Schiavilla, { Phys. Rev. C}, {\bf 51}: 38 (1995)
\bibitem{bigelyas} M. Bigdeli, and S. Elyasi, { Eur. Phys. J. A}, {\bf 51}: 38 (2015)
\bibitem{dh} F. Douchin, and P. Haensel, { Astron. Astrophys}, {\bf 380}: 151 (2001)
\bibitem{byme}  G. Baym, C. Pethick, and D. Sutherland, { ApJ}, {\bf 170}: 299 (1971)
\bibitem{clark} J. W. Clark, and N. C. Chao, { Lettere Nuovo Cimento}, {\bf 2}: 185 (1968)
\bibitem{OBI3} J. C. Owen,  R. F.Bishop, and J. M. Irvine, {Nucl. Phys. A}, {\bf 277}: 45 (1997)
\bibitem{BM98} G. H. Bordbar,and M. Modarres, { Phys. Rev. C}, {\bf 57}: 714 (1998)
\bibitem{MB98}  M. Modarres, and  G. H.Bordbar, { Phys. Rev. C}, {\bf 58}: 2781 (1998)
\bibitem{bordbig6} M. Bigdeli,  G. H. Bordbar, and  A. Poostforush, { Phys. Rev. C}, {\bf 82}: 034309 (2010)
\bibitem{big1} M. Bigdeli, Phys. Rev. C, {\bf82}: 054312 (2010)
{\bibitem{haus} P. E. Haustein, At. Data Nucl. Data Tables, {\bf 39}: 185 (1988)}
\bibitem{ster}  N. Stergioulas, {Living Rev. Rel}, {\bf 6}: 3 (2003)
\bibitem{breu} C. Breu, and  L. Rezzola,   { MNRAS}, {\bf 459}: 646 (2016)

\end{thebibliography}
\end{document}